\documentclass{PoS}
\usepackage{amsmath
,amssymb
,epsfig
,latexsym
,mathrsfs
,slashed
,floatrow
}
\usepackage[font=small,labelfont=bf]{caption}

\newcommand{\fx}{w}

\newcommand{\atildepm}{\overline {{\cal A}} _{i j {\tilde f}_{k}} ^{n + }}
\newcommand{\atildemp}{\overline {{\cal A}} _{i j {\tilde f}_{k}} ^{n - }}

\newcommand{\btildepm}{\overline {{\cal B}} _{i j {\tilde f}_{k}} ^{n + }}
\newcommand{\btildemp}{\overline {{\cal B}} _{i j {\tilde f}_{k}} ^{n - }}

\newcommand{\amub}{a_{\mu;\text{bos}}^{\rm EW(2)}}

\newcommand{\amufrestH}{a_{\mu;\text{f-rest,H}}^{\rm EW(2)}}

\newcommand{\amufrestnoH}{a_{\mu;\text{f-rest,no H}}^{\rm EW(2)}}

\title{Two-loop corrections to $(g-2)_\mu$ \\in the SM and the MSSM}

\ShortTitle{Two-loop corrections to $(g-2)_\mu$ in the SM and the MSSM}

\author{Helvecio\ G. \ Fargnoli
\\
Universidade Federal de Lavras, Lavras, Brazil}
\author{Christoph\ Gnendiger
\\
Institut f\"ur Kern- und Teilchenphysik,
TU Dresden, Dresden, Germany}
\author{Sebastian\ Pa\ss ehr 
\\
Max-Planck Institut f\"ur Physik, M\"unchen, Germany}
\author{Dominik St\"ockinger
\\
Institut f\"ur Kern- und Teilchenphysik,
TU Dresden, Dresden, Germany}
\author{\speaker{Hyejung St\"ockinger-Kim}
\\
Institut f\"ur Kern- und Teilchenphysik,
TU Dresden, Dresden, Germany\\
        E-mail: \email{hyejung.stoeckinger-kim@tu-dresden.de}}

\abstract{
Recent results of two-loop contributions to the muon (g-2) in the Standard Model and its supersymmetric extension are presented. 
In the SM the EW contributions are fixed according to the Higgs boson mass measured at LHC. 
In the MSSM we present the recent result of fermion/sfermion two-loop contributions. The fermion/sfermion contributions are logarithmically enhanced for large sfermion masses and can yield the largest corrections compared to all previously known MSSM two-loop contributions. 
Numerical results in scenarios which are compatible with LHC data are also presented. We find up to 15\% (30\%) corrections for sfermion masses in the 20 TeV (1000 TeV) range.
}

\FullConference{ Loops and Legs in Quantum Field Theory - LL 2014,\\
		 2 April - 2 May 2014 \\
		 Weimar, Germany }

\begin{document}

\section{Introduction}

The muon anomalous magnetic moment, $a_{\mu} = (g-2)_\mu$, is one of the most precise physical measurements and is also a case, where the Standard Model (SM) prediction does not agree with experimental results. 
The deviation between the latest experiment at Brookhaven National Laboratory~\cite{Bennett:2006} and the Standard Model prediction~\cite{SMreviews} amounts to $3...4 \sigma$, and this long-standing deviation motivates new physics. 

It is an ongoing attempt to improve the accuracy of the SM theory prediction. 
From the new physics side the Minimal Supersymmetric Standard Model (MSSM) is one of the most persuasive scenarios to solve this discrepancy. Because of its high precision, $(g-2)_\mu$ is a good tool to constrain parameters of various models. 

The new $(g-2)_\mu$ experiments~\cite{Carey:2009zzb,Roberts:2010cj} are expected to produce results on schedule with much higher accuracy. Challenged by new experiments, the uncertainty in theory predictions should be accordingly reduced in both the SM and the MSSM. 

In these proceedings the recent progress in the SM electroweak (EW) contribution after the Higgs boson mass measurement and the fermion/sfermion two-loop corrections in the MSSM are presented. These new results serve to reduce the uncertainty in the theory predictions. 

\section{The Electroweak SM prediction enhanced by the Higgs boson mass measurement}

\begin{table}[t!]
\begin{tabular}{cccc}
\scalebox{0.22}{\includegraphics[width=\linewidth]{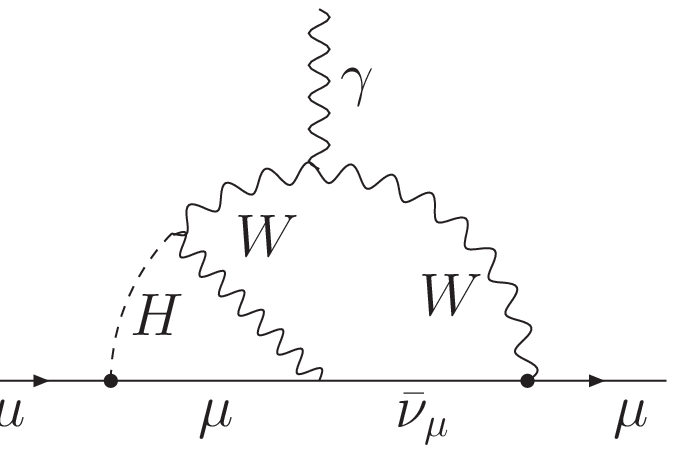}}
&
\scalebox{0.22}{\includegraphics[width=\linewidth]{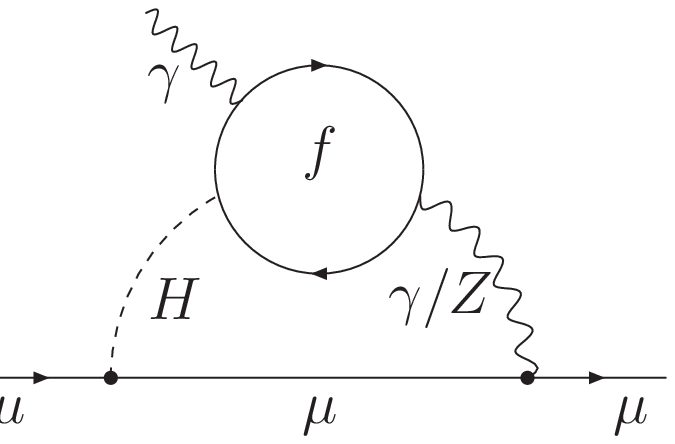}}
& 
\scalebox{0.22}{\includegraphics[width=\linewidth]{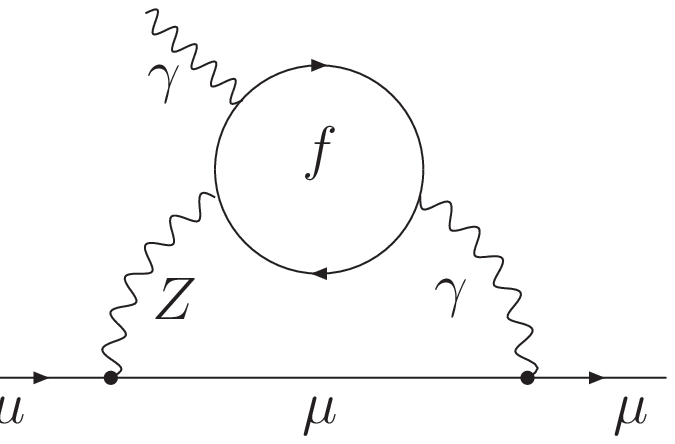}}
& 
\scalebox{0.22}{\includegraphics[width=\linewidth]{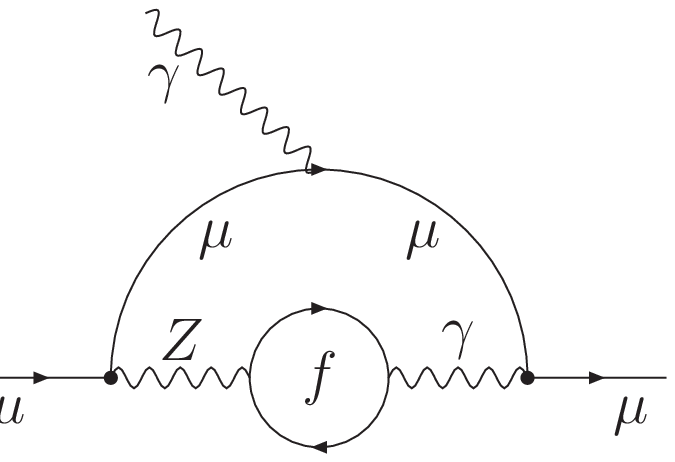}}\\
(a) & (b) & (c) & (d)
\end{tabular}
\captionof{figure}{\label{fig:ew2loop} Four categories of EW two-loop diagrams: Higgs-dependent bosonic (a) and fermionic (b) diagram classes. Diagrams with $\gamma \gamma Z$-fermion loop (c) and with $Z$-$\gamma$ mixing (d).} 
\end{table}

The SM EW one-loop contributions amount to 
$a_\mu^{\rm EW(1)}=(\,194.80\pm 0.01)\times 10^{-11}$. Samples of the EW two-loop Feynman diagrams are shown in Fig.~\ref{fig:ew2loop}. The Higgs dependence is found in the first two categories of diagrams. The most precise estimation of the EW two-loop contributions before the Higgs boson mass measurement was~\cite{CzMV}

\begin{align}
a_{\mu}^{\rm EW} &=  (154 \pm 1 \pm 2) \times 10^{-11},
\label{eq:ew2loop}
\end{align}
where the first error, $\pm 1$, is from the EW hadronic and three-loop contributions, and the second, $\pm 2$, from the Higgs boson mass estimate. 

This earlier estimation has been updated by calculating all Higgs-dependent diagrams exactly and putting the measured Higgs boson mass into the obtained analytic form~\cite{Gnendiger:2013pva}, and thus the uncertainty due to the Higgs boson mass estimate has been eliminated. For the numerical evaluation we employed the Higgs boson mass value $M_H=125.6\pm1.5~{\rm GeV}$, which is an average of the two central values measured by ATLAS and CMS~\cite{ATLASCMS}. The conservative error, $\pm1.5$, covers the $2 \sigma$ range of both measurements.  
The input parameters are the masses of muon, Z-boson and top-quark, the muon decay constant $G_F$, and the fine-structure constant $\alpha$ value at the Thomson limit~\cite{PDG2012}. The W-boson mass is predicted in the SM~\cite{Awramik:2003rn} and we use $M_{\rm W} = 80.363 \pm 0.013~{\rm GeV}$.  

Using these parameters and combining our results with results of Refs.~\cite{CzMV,CKM1,PerisKnecht} we obtain the following EW two-loop contributions:  
\begin{align}
\amub &= (-19.97 \pm 0.03) \times 10^{-11},
\label{eq:ewa}\\
\amufrestH &= (-1.50 \pm 0.01) \times 10^{-11},
\label{eq:ewb}\\
\amufrestnoH &= (-4.64 \pm 0.10) \times 10^{-11},
\label{eq:ewc}\\
a_{\mu}^{\rm EW(2)}(\tau,t,b) &= -(8.21 \pm 0.10) \times 10^{-11},
\label{eq:ewd1}\\
a_{\mu}^{\rm EW(2)}(e,\mu,u,c,d,s) &= -(6.91\pm0.20\pm0.30)\times 10^{-11}.
\label{eq:ewd2} 
\end{align}
Eqs.~(\ref{eq:ewa}) and~(\ref{eq:ewb}) are results with Higgs dependence. 
The largest contribution comes from the bosonic two-loop diagrams, Eq.~(\ref{eq:ewa}), see Fig.~\ref{fig:ew2loop}(a) for a sample diagram. Eq.~(\ref{eq:ewb}) is for Higgs-dependent diagrams with a fermion loop: Fig.~\ref{fig:ew2loop}(b). Eq.~(\ref{eq:ewc}) is the result for $Z-\gamma$ mixing diagrams: Fig.~\ref{fig:ew2loop}(d). Eqs.~(\ref{eq:ewd1}) and~(\ref{eq:ewd2}) are results for diagrams with $\gamma \gamma Z$ interaction with a fermion loop: Fig.~\ref{fig:ew2loop}(c). The former is for the 3rd generation and the latter the 1st and 2nd generation fermions. By combining these two-loop results evaluated with the measured Higgs boson mass and the one-loop result we obtain~\cite{Gnendiger:2013pva}
\begin{align}
a_{\mu}^{\rm EW (2)} &= (153.6 \pm 1.0) \times 10^{-11}, 
\label{eq:ew2result}
\end{align}
where the remaining error, $\pm 1.0$, is due to the electroweak hadronic part and the three and higher order loop contributions. This amount of error is still tolerable after the new experimental result. In Eq.~(\ref{eq:ew2result}) the error due to the uncertainties of $M_H,~m_{t}$ and $M_{\rm W}$ is below $10^{-12}$. 

\begin{figure}[]
\begin{picture}(100,40)(95,120)
\includegraphics[scale=1]{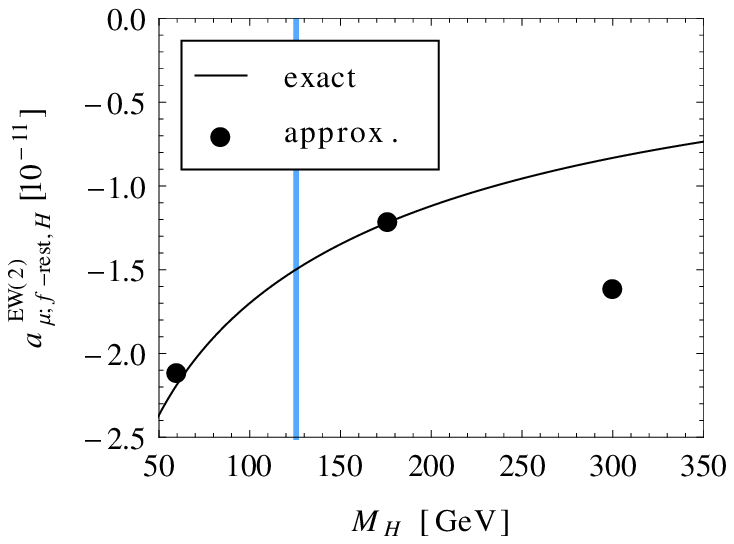}
\put(-105,0){(a)}
\end{picture}
\begin{picture}(200,40)(-20,122)
\includegraphics[scale=1]{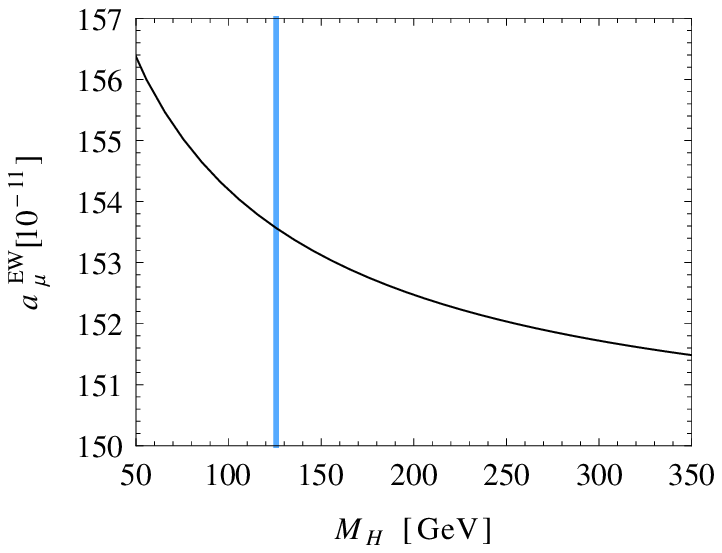}
\put(-105,0){(b)}
\end{picture}
\begin{picture}(300,120)(0,-50)
\captionof{figure}{\label{fig:amufermerstH}
The exact and approximated results for Higgs-dependent diagrams with fermionic loops are compared (a). The numerical result of the EW corrections is shown as a function of the Higgs boson mass (b). In both graphs the vertical blue band indicates the measured Higgs boson mass.}
\end{picture}
\end{figure}

Fig.~\ref{fig:amufermerstH} shows the discrepancy between the exact results~\cite{Gnendiger:2013pva} and the approximated ones~\cite{CKM1}. In the low Higgs boson mass region the exact and approximated results are in agreement, whereas in the large Higgs boson mass region they do not agree with each other. This disagreement originates from the higher order terms of $m_{t}^2/M_{H}^2$, which are neglected in Ref.~\cite{CKM1}. The exact result is, however, important to reduce the error. The second error, $\pm 2$, in Eq.~(\ref{eq:ew2loop}) originated from the three approximated points in Fig.~\ref{fig:amufermerstH} and this error has been eliminated by applying the measured Higgs boson mass value into the exactly calculated analytic result~\cite{Gnendiger:2013pva}.


\section{The MSSM fermion/sfermion two-loop corrections and their non-decoupling behaviour}

MSSM is still one of the most favoured scenarios to explain the $3 \sigma$ deviation, even though many SUSY scenarios with light super particles have already been ruled out by the LHC. 
At this point, it is worth studying non-trivial SUSY mass patterns. 
In Refs.~\cite{shortfsfloops, longfsfloops} several characteristic benchmark points are defined, with which SUSY contributions amount to the current $3 \sigma$ deviation. They involve in particular large mass splittings, for example large $\mu$ (see also Ref.~\cite{Endo}) or heavy left-handed smuon mass.  
It is also relevant to obtain a precise MSSM prediction for $(g-2)_\mu$. 
Here we briefly review the exact fermion/sfermion two-loop results of Refs.~\cite{shortfsfloops, longfsfloops}. 

\begin{figure}[]
\begin{picture}(150,110)(10,5)
\includegraphics[scale=0.8]{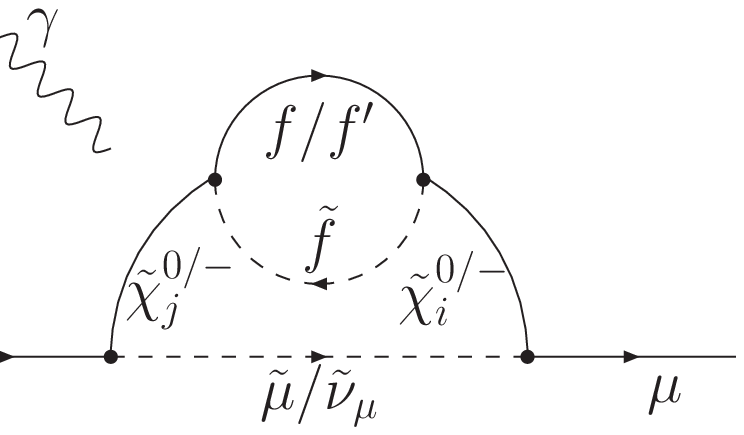}
\put(-105,-45){(a)}
\end{picture}
\begin{picture}(100,50)(-30,-50)
\includegraphics[scale=0.65]{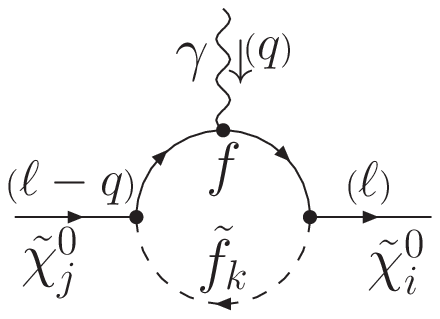}
\end{picture}
\begin{picture}(100,50)(-40,-50)
\includegraphics[scale=0.65]{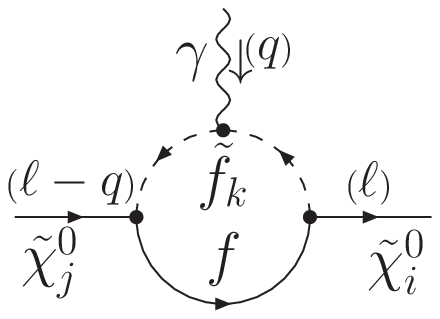}
\end{picture}
\begin{picture}(100,50)(-105,-20)
\includegraphics[scale=0.65]{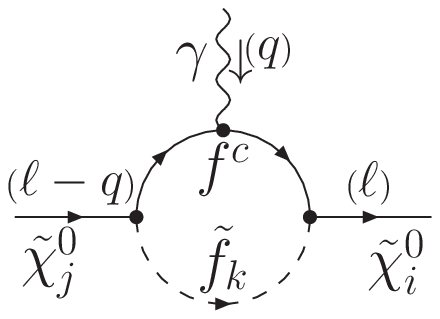}
\end{picture}
\begin{picture}(100,50)(-115,-20)
\includegraphics[scale=0.65]{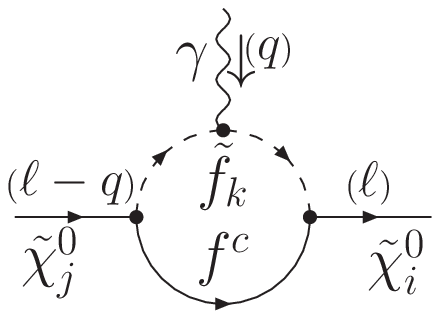}
\put(-105,-20){(b)}
\end{picture}
\begin{picture}(300,10)(0,0)
\captionof{figure}{\label{fig:ngs}The generic fermion/sfermion two-loop Feynman diagram (a) and four inner-loop diagrams contributing to the ${\tilde \chi}^0 \gamma {\tilde \chi}^0$ vertex (b).}
\end{picture}
\end{figure}

The generic fermion/sfermion two-loop Feynman diagram is illustrated in 
Fig.~\ref{fig:ngs}(a). The fermion/sfermion diagrams can be put into four categories according to inside running neutralino or chargino and also the charged particle with which the outer photon couples. The diagrams where the outer photon couples with fermion or sfermion in inner loops are called vertex-type and those where the outer photon couples with smuon or chargino self-energy-type. Using these criteria the fermion/sfermion two-loop diagrams are categorised in four types: neutralino-vertex-type, neutralino-self-energy-type, chargino-vertex-type and chargino-self-energy-type. 

These diagrams do not only have additional parameter dependence on $M_{Ui}$, $M_{Di}$, $M_{Qi}$, $M_{Ei}$ and $M_{Li}$ ($U$, $D$, $Q$, $E$ and $L$ denote the supermultiplet, and $i\,$ denotes the generation) compared to the one-loop and the photonic two-loop contributions~\cite{vonWeitershausen:2010zr} but are also important to investigate the non-conventional mass spectra, where squarks and sleptons have large mass splitting. 

The fermion/sfermion result has yet another important feature: they contain the large universal quantities $\Delta\alpha$ and $\Delta\rho$. The ambiguity in the one-loop results, which is caused by the definition of $\alpha$, is solved by including these fermion/sfermion two-loop contributions with a proper renormalization scheme adapted. 

One method to obtain the two-loop analytic result is the iterated one-loop calculation method: the inner loops are calculated first and their results are inserted into the outer loops to complete the two-loop calculation. The sums of the inner loops produce compact and simple vertices, which makes the calculation effective. For example, the four diagrams in Fig.~\ref{fig:ngs}(b) are the inner-loop diagrams contributing to the neutralino-vertex-type diagrams and the sum of these four diagrams builds an effective ${\tilde \chi}^0 \gamma {\tilde \chi}^0$ vertex, 

\begin{align}
\Gamma _{i j \tilde{f}_{k}} ^{0 \mu} (\ell)=
\frac{1}{16 \pi ^{2}} e Q _{f}\int _{0} ^{1}\frac{{\rm d}\fx}{2} &
\left[ \
\left( \atildepm - \atildemp \gamma ^{5} \right)
\frac{\slashed{\ell}\slashed{q}\gamma^{\mu}-\slashed{\ell}q^{\mu} + \slashed{q}\ell^{\mu}-(\ell\cdot q)\gamma^{\mu}}
{{\cal D}_{f {\tilde f}_{k}}(\ell)}\right.\nonumber\\
&\left.
+\left( \btildepm - \btildemp \gamma ^{5} \right)
\frac{m_{f}}{\fx}\frac{\slashed{q} \gamma ^{\mu} - q^{\mu}}{{\cal D}_{f {\tilde f}_{k}}(\ell) } \
\right], 
\label{eq:ngv}
\end{align}
where $\fx$ is the Feynman parameter, ${\cal D}_{f {\tilde f}_{k}}(\ell) \equiv \ell^2 - m_{f {\tilde f}_{k}}^{2}(\fx)$, with $m_{f {\tilde f}_{k}}^{2}(\fx)\equiv \frac{m_{f}^{2}}{\fx} + \frac{m_{{\tilde f}_{k}}^{2}}{1 - \fx}$. $\atildepm$ and $\btildepm$ are the coupling combinations defined in Ref.~\cite{longfsfloops}. It is not difficult to show that Eq.~(\ref{eq:ngv}) satisfies the Ward-Identity.   

\begin{figure}[]
\begin{picture}(150,170)(10,0)
\includegraphics[scale=0.8]{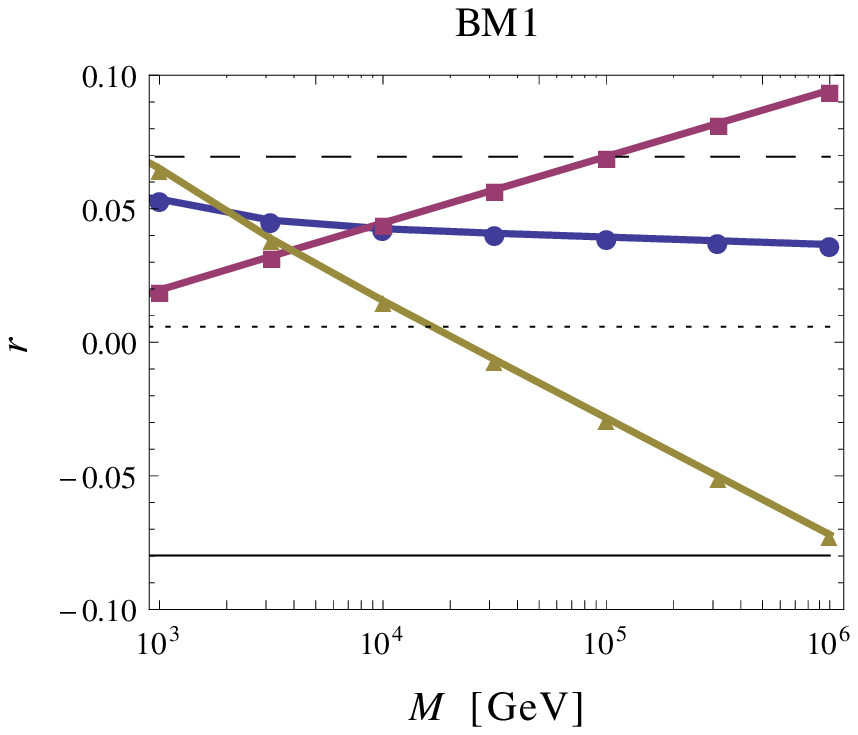}
\put(-95,-13){(a)}
\end{picture}
\begin{picture}(100,50)(0,45)
\includegraphics[scale=0.8]{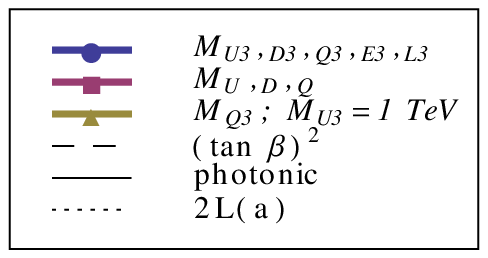}
\end{picture}
\begin{picture}(150,170)(50,0)
\includegraphics[scale=0.8]{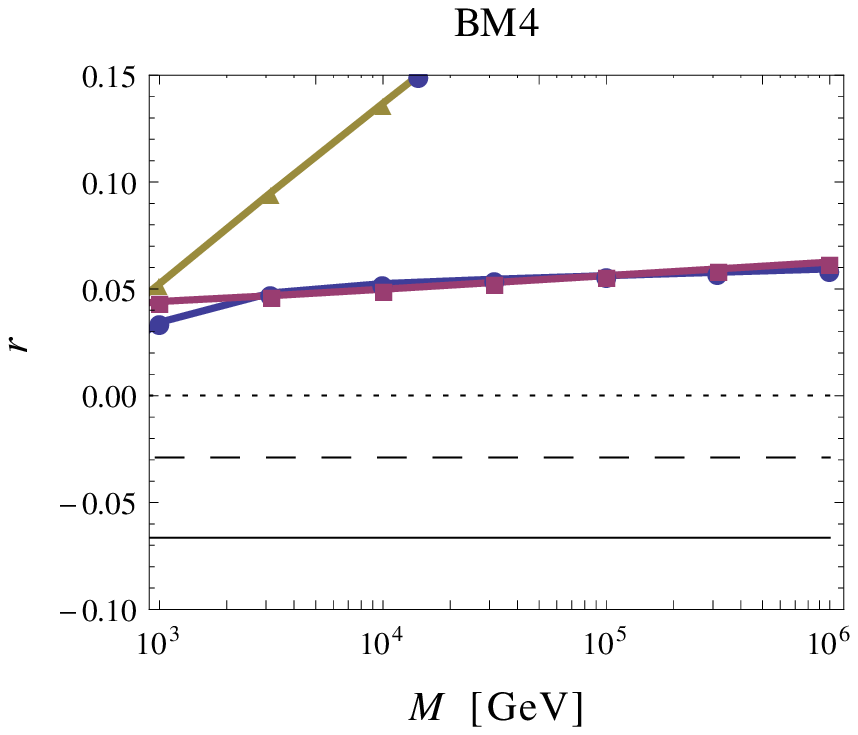}
\put(-95,-13){(b)}
\end{picture}
 \begin{picture}(300,50)(0,-200)
\captionof{figure}{\label{fig:BMs}For BM1 we set $2 M_{1} =  M_{2} = 300~{\rm GeV}$, $m_{{\tilde \mu}_{R}} = m_{{\tilde \mu}_{L}} = 400~{\rm GeV}$, $\mu =350~{\rm GeV}$, and $\tan \beta = 40$, and for BM4 $M_{1} = 140~{\rm GeV}$, $m_{{\tilde \mu}_{R}} = 200~{\rm GeV}$, $M_{2} = m_{{\tilde \mu}_{L}} = 2000~{\rm GeV}$, $\mu = -160~{\rm GeV}$, and $\tan \beta = 40$.}
\end{picture}
\end{figure}

In Fig.~\ref{fig:BMs} the results of numerical analyses with varied squark masses are presented.
Fig.~\ref{fig:BMs}(a) shows the benchmark point BM1 where the smuon masses are of the same order, whereas in BM4 in Fig.~\ref{fig:BMs}(b) there is a large mass splitting between smuon masses. In both Figs.~\ref{fig:BMs}(a) and~\ref{fig:BMs}(b) the non-varied squark masses are kept at $7~{\rm TeV}$ and the third generation slepton masses at $\sim 3~{\rm TeV}$. For the benchmark points in Fig.~\ref{fig:BMs} these new fermion/sfermion two-loop corrections are up to 10\% for small sfermion masses, and up to 30\% for large ones, and their non-decoupling behaviour is observed already at the moderate squark mass scale~\cite{shortfsfloops, longfsfloops}. 
The three additional lines show contributions from photonic two-loop diagrams, $\tan ^2 \beta$-enhanced and $2{\rm L(a)}$ contributions, see Refs.~\cite{vonWeitershausen:2010zr,Marchetti:2008hw,HSW0304}. 
These three lines are constant, since they have essentially no squark mass dependence. 

It is the chargino involving diagrams that are responsible for the non-decoupling behaviour.
The chargino results contain such logarithmic terms as $\ln \frac{m_{\tilde f}}{m_{{\tilde \nu}_{\mu}}}$~\cite{longfsfloops}, and when the mass splitting between $m_{\tilde f}$ and $m_{{\tilde \nu}_{\mu}}$ is large, these logarithmic terms become dominant and the non-decoupling behaviour is observed. On the contrary, when the sfermion masses in the inner loops are of the order of the muon sneutrino mass, the logarithmic terms vanish numerically. 

Depending on the squark and slepton mass scales the fermion/sfermion two-loop corrections range around $(10 ... 30) \%$ to the MSSM one-loop corrections, whereas the photonic two-loop corrections around $-(7 ... 9) \%$~\cite{vonWeitershausen:2010zr}. In the logarithmically enhanced parameter range it is possible to use the leading logarithmic approximation. A very compact approximation formula is found in Ref.~\cite{longfsfloops} and its Mathematica implementation in Ref.~\cite{code}, which serves as a useful tool to investigate various mass spectra.


\begin{thebibliography}{99}

\bibitem{Bennett:2006}G.W. Bennett, et al.,
(Muon $(g-2)$ Collaboration), Phys. Rev. D {\bf 73}, 072003 (2006).

\bibitem{SMreviews}
  F.~Jegerlehner and A.~Nyffeler,
  Phys.\ Rept.\  {\bf 477} (2009) 1;
  J.~P.~Miller, E.~d.~Rafael, B.~L.~Roberts and D.~St\"ockinger,
  Ann.\ Rev.\ Nucl.\ Part.\ Sci.\  {\bf 62} (2012) 237.
\bibitem{Carey:2009zzb}
  R.~M.~Carey, K.~R.~Lynch, J.~P.~Miller, B.~L.~Roberts, W.~M.~Morse, Y.~K.~Semertzides, V.~P.~Druzhinin and B.~I.~Khazin {\it et al.},
  FERMILAB-PROPOSAL-0989.
\bibitem{Roberts:2010cj}
  B.~L.~Roberts,
  Chin.\ Phys.\ C {\bf 34} (2010) 741
  [arXiv:1001.2898 [hep-ex]].
\bibitem{CzMV}
  A. Czarnecki, W.~J. Marciano, A. Vainshtein
  Phys.Rev.D {\bf 67} (2003) 073006, Erratum-ibid.D{\bf 73} (2006)
  119901.
\bibitem{CKM1}
  A.~Czarnecki, B.~Krause and W.~J.~Marciano,
  Phys.\ Rev.\ D {\bf 52} (1995) 2619.
%
\bibitem{PerisKnecht}
  S.~Peris, M.~Perrottet and E.~de Rafael,
  Phys.\ Lett.\ B {\bf 355} (1995) 523;
  M.~Knecht, S.~Peris, M.~Perrottet and E.~De Rafael,
  JHEP {\bf 0211} (2002) 003.
\bibitem{Gnendiger:2013pva}
  C.~Gnendiger, D.~St\"ockinger and H.~St\"ockinger-Kim,
  Phys.\ Rev.\ D {\bf 88} (2013) 053005.
\bibitem{ATLASCMS}
  [ATLAS Collaboration],
  ATLAS-CONF-2013-014;
%
  [CMS Collaboration],
  CMS-PAS-HIG-13-005.
%
\bibitem{PDG2012}
  J.~Beringer et al. (Particle Data Group)
  Phys.\ Rev.\ D {\bf 86} (2012) 010001.
\bibitem{Awramik:2003rn}
  M.~Awramik, M.~Czakon, A.~Freitas and G.~Weiglein
  Phys.\ Rev.\ D {\bf 69} (2004) 053006.
\bibitem{shortfsfloops}
  H.~Fargnoli, C.~Gnendiger, S.~Passehr, D.~St\"ockinger and H.~St\"ockinger-Kim,
  arXiv:1309.0980 [hep-ph].
\bibitem{longfsfloops}
  H.~Fargnoli, C.~Gnendiger, S.~Passehr, D.~St\"ockinger and H.~St\"ockinger-Kim,
  arXiv:1311.1775 [hep-ph].
\bibitem{Endo}
  M.~Endo, K.~Hamaguchi, S.~Iwamoto and T.~Yoshinaga,
  arXiv:1303.4256 [hep-ph];
  M.~Endo, K.~Hamaguchi, T.~Kitahara and T.~Yoshinaga,
  arXiv:1309.3065 [hep-ph];
  M.~Endo, K.~Hamaguchi, S.~Iwamoto, T.~Kitahara and T.~Moroi,
  arXiv:1310.4496 [hep-ph].
\bibitem{vonWeitershausen:2010zr}
  P.~von Weitershausen, M.~Sch\"afer, H.~St\"ockinger-Kim and D.~St\"ockinger,
  Phys.\ Rev.\ D {\bf 81} (2010) 093004.
\bibitem{Marchetti:2008hw}
  S.~Marchetti, S.~Mertens, U.~Nierste and D.~St{\"o}ckinger,
  Phys.\ Rev.\  D {\bf 79}, 013010 (2009).
\bibitem{HSW0304}
  S.~Heinemeyer, D.~St\"ockinger and G.~Weiglein,
  Nucl.\ Phys.\ B {\bf 690} (2004) 62;
  Nucl.\ Phys.\ B {\bf 699} (2004) 103.
%
\bibitem{code}
The Mathematica implementation 
  of the approximation formula can be
  obtained from 
  {\tt http://iktp.tu-dresden.de/?id=theory-software}.

\end{thebibliography}
\end{document}